\documentclass[preprint,11pt,authoryear]{elsarticle}

\usepackage[T1]{fontenc}
\usepackage{lmodern}
\usepackage{amsmath,amssymb,amsthm}
\usepackage{graphicx}
\usepackage{booktabs}
\usepackage{microtype}
\usepackage{xcolor}
\usepackage{hyperref}
\usepackage[capitalise]{cleveref}
\usepackage{bm}

\hypersetup{
  colorlinks=true,
  linkcolor=black!75!blue,
  citecolor=black!75!blue,
  urlcolor=black!50!blue
}

\journal{Physica A: Statistical Mechanics and its Applications}

\theoremstyle{plain}
\newtheorem{theorem}{Theorem}
\newtheorem{lemma}[theorem]{Lemma}
\newtheorem{proposition}[theorem]{Proposition}
\newtheorem{corollary}[theorem]{Corollary}
\theoremstyle{definition}
\newtheorem{definition}{Definition}
\theoremstyle{remark}
\newtheorem{remark}[theorem]{Remark}

\newcommand{\Lp}{L^{+}}
\newcommand{\Tp}{\tau^{+}}
\newcommand{\HVG}{\mathrm{HVG}}
\newcommand{\E}{\mathbb{E}}
\newcommand{\Prob}{\mathbb{P}}
\newcommand{\Var}{\mathrm{Var}}
\newcommand{\R}{\mathbb{R}}
\newcommand{\fbm}{B^{H}}

\begin{document}

\begin{frontmatter}

\title{First-passage horizons in horizontal visibility graphs: a rank-invariant estimator of path roughness for rough volatility models}

\author[uw]{Micha{\l} Sikorski\corref{cor1}\fnref{fn1}}
\ead{mm.sikorski3@student.uw.edu.pl}
\address[uw]{Interdisciplinary Centre for Mathematical and Computational Modelling, University of Warsaw, Warsaw, Poland}
\cortext[cor1]{Corresponding author}
\fntext[fn1]{ORCID: \href{https://orcid.org/0009-0009-4253-9489}{0009-0009-4253-9489}}

\begin{abstract}
Horizontal visibility graphs (HVGs) encode the ordinal structure
of time series and provide graph-local summaries of path topology.
This article introduces $\Lp(t)$, the forward visibility horizon
at node $t$, with finite-sample terminal non-crossings treated as
right-censored observations. For paths without ties, each
uncensored $\Lp(t)$ is identical to the first-passage time
$\Tp(t)=\inf\{k\ge1:x_{t+k}\ge x_t\}$. For an i.i.d.\ sequence
with a continuous distribution, the survival law is exactly
$\Pr[\Lp\ge k]=1/k$, equivalent to R\'enyi's record statistic and
implying infinite mean and variance. Hence roughness is estimated
on a power-law survival scale through a single tail exponent
$\theta$. Combining the identity $\Lp=\Tp$ with discrete-grid
persistence theory for fractional Brownian motion gives the
prediction $\theta(H)=1-H$. For rough Bergomi-type volatility,
the same prediction is derived under an explicit persistence
hypothesis for Riemann--Liouville fBm increments and verified
numerically. In Monte-Carlo experiments ($N=10{,}000$, $T=2^{16}$), a
Hill-MLE with Clauset--Shalizi--Newman threshold selection
recovers $\theta(H)$ within one cross-replicate standard deviation
for $H\le0.2$ and reveals a positive finite-size bias for smoother
paths. The rank-invariant, parameter-free estimator
separates rough Bergomi volatility from classical Heston, GARCH,
and FIGARCH benchmarks. Applied to daily
FRED VIX data from 2000--2026, the rolling estimate is
$\hat\theta=0.91\pm0.19$ across 45 four-year windows and lies far
below an overlapping-window i.i.d.\ Monte-Carlo null ($p<0.001$).
The statistic offers an ordinal diagnostic of
roughness for financial volatility and other complex time-series
systems.
\end{abstract}


\end{frontmatter}

\section{Introduction}\label{sec:intro}

Complex systems are often observed through scalar time series whose
amplitudes are noisy, model-dependent, or not directly comparable
across systems, while their ordinal geometry carries robust
information about intermittency, persistence, and path roughness.
This is true in financial markets, but also in many statistical
physics settings where one wants to compare trajectories generated
by different microscopic mechanisms through a common topological
summary. A rank-invariant graph statistic is attractive in this
setting because it depends only on the order structure of the path,
not on the marginal distribution or on a fitted parametric scale.

The estimation of the Hurst index $H$ of a volatility process is a
particularly demanding instance of this broader problem and a
natural benchmark for a roughness diagnostic. Since the empirical
discovery that log-realised variance behaves as fractional Brownian
motion (fBm) with $H \approx 0.1$ on a wide range of underlying
indices and time scales \citep{gjr2014, bayerfrizgatheral2016},
roughness estimation has become a fundamental problem in financial
econometrics and econophysics. Standard estimators---detrended
fluctuation analysis (DFA), Whittle, wavelet variance, R/S---require
strong stationarity assumptions or make specific hypotheses about
the noise structure of the observable. They also confound
long-memory effects with path roughness when applied to time series
that exhibit both \citep{fukasawa2022}.

The horizontal visibility graph (HVG) provides an alternative,
purely combinatorial encoding of the geometry of a path
\citep{lacasa2008, luque2009}. The HVG of a length-$T$ sequence
$x = (x_1, \ldots, x_T)$ is the simple graph on the vertex set
$\{1, \ldots, T\}$ in which two indices $i < j$ are connected if
and only if every intermediate value satisfies
$x_m < \min(x_i, x_j)$. Properties of HVGs have been used as
diagnostics of irreversibility, chaos, and long memory in time
series \citep{lacasa2012kld, lacasa2009fbm, iacovacci2017}, but
existing analytical results focus on the degree distribution and
on Kullback--Leibler-type asymmetries. The forward-horizon
statistic $\Lp(t)$, defined below, has not, to our knowledge,
been studied as an estimator of path roughness, despite being a
natural local object on the graph.

This paper makes six contributions: it (i) proves that each
uncensored $\Lp(t)$ coincides
with the first-passage time
$\Tp(t) = \inf\{k\ge 1 : x_{t+k} \ge x_t\}$,
(ii) treats terminal non-crossings in finite samples as
right-censored observations, (iii) derives the closed-form i.i.d.\ null
$\Pr[\Lp \ge k] = 1/k$, (iv) combines the latter with Molchan's
persistence theorem for fBm to obtain $\theta(H) = 1 - H$, (v)
constructs a Hill-MLE estimator with data-driven threshold and
block-bootstrap confidence band, and (vi) validates it on fBm,
Bergomi-type rough volatility, classical Heston, GARCH(1,1) and
FIGARCH(1,$d$,1). The paper closes with a deployment on twenty-six years
of daily CBOE VIX data, where the rolling estimator returns
values consistent with the rough-volatility regime
$H_{\mathrm{vol}} \approx 0.1$.

\paragraph{The novelty} The fact that fBm has persistence
exponent $1 - H$ has been known since \citet{molchan1999}; the
fact that the HVG carries a degree distribution that depends on
the underlying process is the subject of an extensive
econophysics literature \citep{lacasa2008, luque2009}. The
contribution here is that a purely ordinal HVG statistic---the
uncensored forward visibility horizon $\Lp$---is \emph{exactly} a
first-passage statistic and therefore inherits the persistence
exponent. Finite-sample horizons with no observed crossing before
the right boundary are treated as right-censored; the main estimator
uses a complete-case Hill fit, with censored-likelihood and
lower-bound sensitivity checks reported below. This yields a rank-invariant,
parameter-free roughness diagnostic with a closed-form i.i.d.\ null,
derived from a single combinatorial identity (Lemma~\ref{thm:Lp})
rather than from a fitted distribution or a moment estimator.

\paragraph{Relation to existing HVG Hurst estimators}
Earlier visibility-graph Hurst estimators use global
degree-spectrum information and typically require an empirical
calibration between a graph-spectrum parameter and $H$
\citep{lacasa2009fbm}. The statistic proposed here instead uses a local horizon
variable, one $\Lp(t)$ per node, whose i.i.d.\ survival law is exact
and whose fBm exponent follows from first-passage persistence. The
point is therefore not that degree-based graph summaries are
non-ordinal, but that the horizon statistic has a closed-form null
and a direct persistence interpretation rather than only an
empirical calibration curve.

\section{Forward visibility horizons and the i.i.d.\ null}
\label{sec:thy}

\begin{definition}[Forward visibility horizon]
\label{def:horiz}
Let $x = (x_1, \ldots, x_T) \in \R^T$ be a path with no ties.
The \emph{forward visibility horizon} at node $t$ is the
HVG-visible endpoint that first reaches or exceeds the anchor,
\[
  \Lp(t) \;=\; \sup\bigl\{k \ge 1 : (t, t+k) \in E[\HVG(x)],
                            \; x_{t+k} \ge x_t,\; t+k \le T\bigr\}.
\]
If this set is empty on a finite path, the observation at $t$ is
right-censored rather than assigned a numerical value. Here
$E[\HVG(x)]$ is the edge set of the horizontal visibility graph
of $x$.
\end{definition}

\begin{figure}[t]
  \centering
  \includegraphics[width=\linewidth]{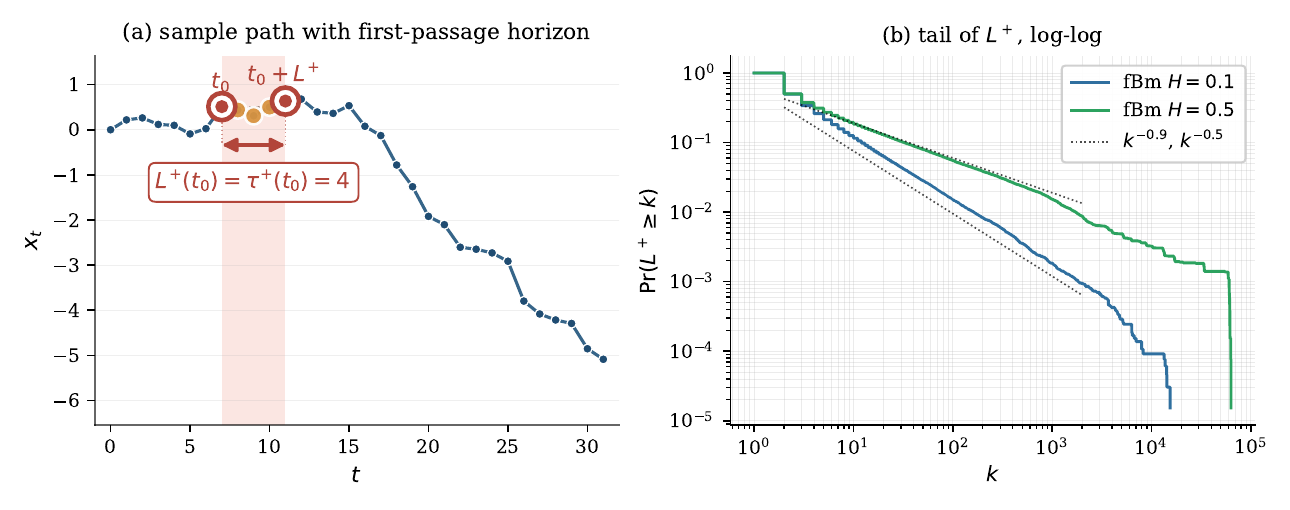}
  \caption{(a) Sample path of length $T = 32$; the highlighted
  node $t_0$ has $\Lp(t_0) = \Tp(t_0)$. (b) Empirical survival of
  uncensored $\Lp$ for two fBm paths of length $T = 2^{16}$. Reference
  Pareto slopes $k^{-0.9}$ and $k^{-0.5}$ correspond respectively
  to $H = 0.1$ and $H = 0.5$ via the prediction $\theta = 1 - H$.}
  \label{fig:schem}
\end{figure}

Figure~\ref{fig:schem}(a) illustrates a representative path with
its $\Lp(t_0)$. Figure~\ref{fig:schem}(b) shows the empirical
survival distribution of uncensored $\Lp$ for two fBm samples; the
asymptotic slope on a log--log scale is the tail exponent
$\theta$ estimated below.

\begin{lemma}[$\Lp = \Tp$ for uncensored horizons]
\label{thm:Lp}
For any path with no ties, if there exists $k \ge 1$ such that
$t+k \le T$ and $x_{t+k} \ge x_t$, then $\Lp(t) = \Tp(t)$ where
\[
  \Tp(t) \;=\; \inf\{k \ge 1 : x_{t+k} \ge x_t\},
\]
and terminal observations for which this set is empty are
right-censored. On an infinite i.i.d.\ path the latter event has
probability zero.
\end{lemma}

\begin{proof}
Set $k^{*} = \Tp(t)$ and assume that it is observed before the
right boundary.
For all $1 \le m < k^{*}$ one has $x_{t+m} < x_t$ by definition
of $\Tp$, hence $x_{t+m} < \min(x_t, x_{t+k^{*}})$. The HVG
visibility condition is satisfied, so
$(t, t+k^{*}) \in E[\HVG(x)]$, giving $\Lp(t) \ge k^{*}$.
Conversely, for any $k > k^{*}$, the index $m = k^{*}$ is
intermediate. Visibility from $t$ to $t+k$ would require
$x_{t+k^{*}} < \min(x_t, x_{t+k})$, but the definition of $\Tp$
gives $x_{t+k^{*}} \ge x_t \ge \min(x_t, x_{t+k})$, a
contradiction. Hence $\Lp(t) = k^{*}$.
\end{proof}

After Lemma~\ref{thm:Lp}, the HVG representation is not needed
computationally: the statistic can be computed by the
next-greater-or-equal problem. Its role is conceptual. The lemma
identifies a graph-local, ordinal observable whose survival tail
has first-passage meaning, thereby linking visibility-graph
diagnostics with persistence theory and explaining why
horizon-based HVG statistics should be analysed on a power-law
survival scale rather than by finite moments.

\begin{proposition}[i.i.d.\ null]
\label{thm:iid}
Let $X_1, X_2, \ldots$ be i.i.d.\ with a common continuous
distribution. Then for every integer $k \ge 1$,
\[
  \Prob\!\left[\Lp(t) \ge k\right] \;=\; \frac{1}{k},
\]
and $\E[\Lp(t)] = \infty$, $\Var[\Lp(t)] = \infty$.
\end{proposition}

\begin{proof}
By Lemma~\ref{thm:Lp}, $\Lp(t) \ge k$ iff
$X_{t+1}, \ldots, X_{t+k-1} < X_t$, i.e.\ iff $X_t$ is the
maximum of the i.i.d.\ block
$\{X_t, X_{t+1}, \ldots, X_{t+k-1}\}$ of size $k$. By
exchangeability of ranks under a continuous distribution, each of
the $k$ positions is equally likely to host the maximum, so the
probability equals $1/k$. This is the classical record statistic
of \citet{renyi1962}; see also
\citet[Ch.~2.2]{arnoldbalakrishnannagaraja1998}. The series
$\sum_{k\ge 1} \Pr(\Lp \ge k) = \sum_{k \ge 1} 1/k$ is the
harmonic series and diverges; hence $\E[\Lp] = \infty$. The
divergence of the variance follows from
$\E[(\Lp)^2] = \sum_{k\ge 1}(2k-1)\Pr(\Lp\ge k)$.
\end{proof}

Proposition~\ref{thm:iid} provides the closed-form null model on
a power-law scale: under independence, the survival exponent of
$\Lp$ is exactly $\theta_{\mathrm{iid}} = 1$. The estimator
constructed in Section~\ref{sec:est} therefore measures a tail
exponent rather than a finite moment; this is necessary because
the divergence of $\E[\Lp]$ implies that any unbiased estimator
of the mean horizon must itself diverge under the i.i.d.\ null.
Working with the tail exponent yields a single number
$\theta \in [0, \infty)$ on which path roughness is monotonically
encoded.

\section{Estimator}\label{sec:est}

Given a sample path $x$ of length $T$, $\Lp(t)$ is computed for
$t = 1, \ldots, T$ in $\mathcal{O}(T)$ time using a monotone stack
that solves the next-greater-or-equal problem. Only
indices for which $\Tp(t)$ is observed before the right boundary;
terminal non-crossings are treated as right-censored. The main
Hill fit is a complete-case estimate on uncensored horizons, giving
a sample
$\bm{L} = (\Lp(t_i))_{i=1}^{n}$ with $n \le T-1$.
For empirical series with ties, the implementation uses deterministic time-order
tie breaking: if $x_{t+k}=x_t$ and $k>0$, the later value is treated
as the first crossing. On VIX this convention is numerically
immaterial: replacing it by the stricter convention that equal
future values do not cross changes the rolling mean by $0.003$.

The Hill-MLE estimator with threshold $\ell_0 \ge 1$ is
\begin{equation}
  \hat\theta(\ell_0) \;=\; \frac{n_{\ell_0}}{\sum_{L_i \ge \ell_0}
  \log(L_i/\ell_0)}, \qquad
  n_{\ell_0} = \#\{i : L_i \ge \ell_0\},
\label{eq:hill}
\end{equation}
which is the maximum-likelihood estimator under a continuous
Pareto tail and admits the asymptotic standard error
$\hat\theta/\sqrt{n_{\ell_0}}$ for i.i.d.\ tail samples
\citep{hill1975}. Because $\bm{L}$ is dependent across $t$ (a
visibility horizon involves overlapping windows), the
asymptotic SE is replaced by a stationary block bootstrap of
\citet{politisromano1994} with mean block length
$b = \lceil T^{1/3} \rceil$ and $B = 200$ resamples; the
dependence-aware confidence intervals are reported throughout.

The threshold $\ell_0$ is selected by the protocol of
\citet{clausetshalizinewman2009}: for each candidate $\ell_0$ on
a geometric grid, $\hat\theta(\ell_0)$ is computed via
Eq.~\eqref{eq:hill} and the Kolmogorov--Smirnov distance between
the empirical tail of $\bm{L}$ above $\ell_0$ and the
continuous-Pareto reference with exponent $\hat\theta(\ell_0)$.
The minimiser is reported as $\ell_0^{\star}$.
In the computations, the candidate set contains all integer
thresholds from $1$ to $\min(19,\lfloor q_{0.9}\rfloor)$ and
30 logarithmically spaced integer thresholds between $20$ and the
empirical $90\%$ quantile $q_{0.9}$, with duplicates removed.
Alternative tail-index estimators, such as Pickands or
bias-corrected Hill variants, could be substituted; the Hill-MLE
is used here because it is the likelihood estimator for the
Pareto survival model and extends directly to the censored-tail
check below.
Figure~\ref{fig:appendix}(b) shows that the resulting Hill plot
is essentially flat in a wide neighbourhood of $\ell_0^{\star}$.

As a finite-sample check on right censoring, two
non-headline alternatives are computed at the selected
$\ell_0^{\star}$. Let
$C_j$ be the lower bound for a right-censored terminal horizon.
The censored Pareto likelihood gives
\[
  \hat\theta_{\mathrm{cens}}
  = \frac{n_u}{
      \sum_{i\in u}\log(L_i/\ell_0^{\star})
      + \sum_{j\in c}\log(C_j/\ell_0^{\star})},
\]
where $u$ and $c$ denote uncensored and censored tail observations.
A conservative lower-bound fit is also reported; it treats each
censored $C_j$ as if it were the exact horizon. These diagnostics
are used only to assess sensitivity to terminal censoring.

\section{Validation on fractional Brownian motion}
\label{sec:fbm}

\subsection{Predicted exponent}
Let $\fbm$ be standard fBm of Hurst index $H \in (0, 1)$, sampled
on the integer grid. By Lemma~\ref{thm:Lp},
$\Lp(t) = \Tp(t) = \inf\{k \ge 1 : \fbm_{t+k} \ge \fbm_t\}$.
Because the estimator is applied to integer-grid samples, the
relevant persistence probability is the discrete one
\begin{equation}
  p_k(H) \;=\;
  \Prob\!\left[\fbm_1 < 0,\ldots,\fbm_k < 0\right]
  \;=\; k^{-(1-H)+o(1)} \quad (k \to \infty).
\label{eq:molchan}
\end{equation}
This exponent is the integer-grid persistence counterpart of
\citet{molchan1999}, usually formulated through a fixed positive
barrier such as $\Prob[\sup_{0\le s\le T}\fbm_s \le 1]$; see also
the persistence survey of \citet{aurzadasimon2015}. The event
$\{\Tp(t) > k\}$ is equivalent to
$\{\fbm_{t+j} - \fbm_t < 0,\; j=1,\ldots,k\}$, which by
stationarity of increments has probability $p_k(H)$. Hence:

\begin{corollary}
\label{cor:molchan}
For fBm with Hurst index $H$,
$\Prob[\Lp(t) \ge k] = k^{-(1-H)+o(1)}$, equivalently the
survival-tail exponent is
$\theta(H) = 1 - H$.
\end{corollary}

Corollary~\ref{cor:molchan} is a consequence of
Lemma~\ref{thm:Lp} combined with \citet{molchan1999} and is not a
new claim of the persistence literature; it serves as a
falsifiable prediction for the estimator of Section~\ref{sec:est}.

\paragraph{Intuition} For small $H$ paths are rough, with
frequent local extrema, and the first-passage time $\Tp$ has a
thin tail (large $\theta$). For large $H$ paths are smooth and
exhibit long monotone runs, so $\Tp$ has a heavy tail (small
$\theta$). Corollary~\ref{cor:molchan} quantifies this as an
exact linear relation $\theta = 1 - H$.

\subsection{Numerical validation}
\begin{figure}[t]
  \centering
  \includegraphics[width=\linewidth]{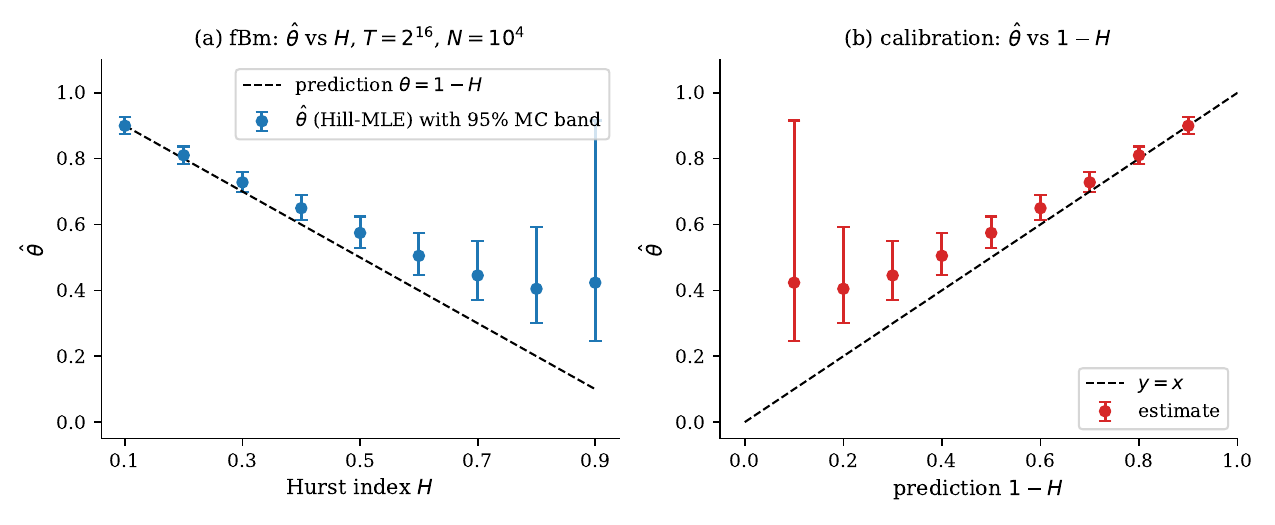}
  \caption{(a) $\hat\theta(H)$ for fBm with Davies--Harte
  sampling, $T = 2^{16}$, $N = 10{,}000$. Bars are Monte-Carlo
  $95\%$ envelopes. (b) Calibration plot $\hat\theta$ versus the
  predicted $1 - H$.}
  \label{fig:fbm}
\end{figure}

The validation generates $N = 10{,}000$ fBm paths of length
$T = 2^{16}$ for each
$H \in \{0.1, 0.2, \ldots, 0.9\}$ via the Davies--Harte algorithm
\citep{daviesharte1987} and applies the estimator of
Section~\ref{sec:est}. Figure~\ref{fig:fbm} reports point
estimates and Monte-Carlo $95\%$ bands across replicates.

Table~\ref{tab:fbm} lists the numerical results. The estimator
recovers $\theta(H)$ within one cross-replicate standard deviation
for $H \le 0.2$. From $H = 0.3$ upward the large-$N$ run resolves
a positive finite-size bias that grows with $H$: this is the regime
of very heavy survival tails
($\theta = 1-H$ close to zero), in which the rare events that
dominate the Hill estimator are scarce at the available $T$.
Figure~\ref{fig:appendix}(a) shows that this bias decays
monotonically with $T$; for $H = 0.1$ it is below $0.01$ at
$T = 2^{18}$.

\begin{table}[t]
\centering
\small
\caption{fBm Davies--Harte ensemble at $T = 2^{16}$, $N = 10{,}000$.
$\hat\theta$ is the cross-replicate mean of the Hill-MLE with CSN
threshold; $\sigma$ is the cross-replicate standard deviation.
Predicted $\theta = 1 - H$.}
\label{tab:fbm}
\begin{tabular}{@{}lccccc@{}}
\toprule
$H$ & $\hat\theta$ & $\sigma$ & predicted & bias &
$\bar\ell_0^{\star}$ \\
\midrule
0.1 & 0.900 & 0.013 & 0.90 & $-0.0003$ & 20 \\
0.2 & 0.810 & 0.013 & 0.80 & $+0.010$ & 20 \\
0.3 & 0.728 & 0.015 & 0.70 & $+0.028$ & 20 \\
0.4 & 0.650 & 0.019 & 0.60 & $+0.050$ & 20 \\
0.5 & 0.575 & 0.024 & 0.50 & $+0.075$ & 19 \\
0.6 & 0.505 & 0.032 & 0.40 & $+0.105$ & 14 \\
0.7 & 0.445 & 0.046 & 0.30 & $+0.145$ & 9 \\
0.8 & 0.405 & 0.075 & 0.20 & $+0.205$ & 7 \\
0.9 & 0.423 & 0.183 & 0.10 & $+0.323$ & 15 \\
\bottomrule
\end{tabular}
\end{table}

\subsection{Control processes}
Table~\ref{tab:ctrl} reports $\hat\theta$ on processes whose
answer is known a priori at $T = 2^{16}$. The i.i.d.\ baseline in
Table~\ref{tab:ctrl} (rows 1--2) lies $\sim 0.02$ above the
asymptotic value $\theta = 1$ of Proposition~\ref{thm:iid}; this
is the standard finite-sample bias of the Hill-MLE under a
discrete tail. It decays monotonically with $T$: a separate run
on i.i.d.\ uniform yields
$\hat\theta = 1.041, 1.024, 1.016, 1.014$ at
$T = 2^{14}, 2^{16}, 2^{18}, 2^{20}$ respectively ($N = 50$
replicates each), the second of which agrees with the value
reported in Table~\ref{tab:ctrl}. Brownian motion gives the fBm
$H = 0.5$ value. GBM gives the same estimate as Brownian motion:
this confirms rank-invariance, since geometric Brownian motion is
a strictly increasing transformation of arithmetic Brownian
motion applied uniformly across $t$. The OU process at fast and
slow sampling resolves a smooth crossover: at dense sampling
($\Delta t = 0.01$) the path is locally Brownian, giving a value
between $\theta_{\mathrm{BM}}$ and $\theta_{\mathrm{iid}}$; at
sparse sampling ($\Delta t = 1.0$) consecutive observations are
nearly independent and the estimator approaches the i.i.d.\
value.

\begin{table}[t]
\centering
\small
\caption{Control processes, $T = 2^{16}$, $N = 50$.
SE = $\sigma / \sqrt{N}$.}
\label{tab:ctrl}
\begin{tabular}{@{}lcccc@{}}
\toprule
process & $\hat\theta$ & SE & $\sigma$ & predicted \\
\midrule
i.i.d.\ uniform & 1.021 & 0.002 & 0.012 &
1.00 (Prop.~\ref{thm:iid}) \\
i.i.d.\ Gaussian & 1.024 & 0.002 & 0.014 & 1.00 \\
Brownian motion & 0.569 & 0.004 & 0.025 & 0.50 (fBm $H{=}0.5$) \\
OU, $\Delta t = 0.01$ & 0.653 & 0.002 & 0.014 & --- (locally BM) \\
OU, $\Delta t = 1.0$ & 1.052 & 0.002 & 0.011 & 1.00 (sparse) \\
GBM, $\sigma=0.2$, $\mu=0.05$ & 0.570 & 0.004 & 0.025 &
0.50 (rank-inv.) \\
\bottomrule
\end{tabular}
\end{table}

\section{Application to rough volatility}\label{sec:roughvol}

The application uses the rough Bergomi-type variance model of
\citet{bayerfrizgatheral2016},
\begin{equation}
  v_t \;=\; \xi_0 \exp\!\left(\eta\, Z^H_t
        \;-\; \tfrac{1}{2}\eta^2 t^{2H}\right),
  \label{eq:rb}
\end{equation}
where $Z^H$ is the Riemann--Liouville fBm
$Z^H_t = \sqrt{2H}\!\int_0^t (t-s)^{H-1/2} dW_s$, simulated by
the hybrid scheme of \citet{bennedsenlundepakkanen2017} with
$\kappa = 1$. The estimator is applied to the variance
trajectory $v_t$. Note that, in contrast to the case where
rank-invariance applies trivially, the time-dependent drift
$-\tfrac{1}{2}\eta^2 t^{2H}$ in Eq.~\eqref{eq:rb} prevents a
uniform monotone identification of $v_t$ with $Z^H_t$: the map
$z \mapsto v_t(z) = \xi_0 \exp(\eta z - \tfrac{1}{2}\eta^2
t^{2H})$ varies with $t$. Rank-invariance still gives
$\Lp(v) = \Lp(\log v)$, since $\log$ is applied uniformly across
$t$, but $\Lp(\log v) \neq \Lp(Z^H)$ in general.
Corollary~\ref{cor:molchan} therefore does not transfer
automatically; the corresponding prediction is derived below under
an explicit persistence hypothesis for RL-fBm increments.

\subsection{Asymptotic exponent for rough Bergomi}
\label{sec:rb-theory}

By rank-invariance under the uniform monotone transformation
$\log$, $\Lp$ on $v$ equals $\Lp$ on $\log v$. The first-passage
event for $\log v$ rearranges, after subtracting common terms,
into
\begin{equation}
  \{\Tp(t) > k\} \;=\;
  \Bigl\{\,Z^H_{t+j} - Z^H_t \;<\; \tfrac{\eta}{2}\bigl[(t+j)^{2H}
    - t^{2H}\bigr]
    \quad j=1,\ldots,k\Bigr\},
  \label{eq:rb-event}
\end{equation}
i.e.\ the persistence of the RL-fBm increment process
$j \mapsto Z^H_{t+j} - Z^H_t$ below the deterministic threshold
$g_t(s) = \tfrac{\eta}{2}[(t+s)^{2H} - t^{2H}]$.

\begin{proposition}[Conditional persistence result for rough
Bergomi-type volatility]
\label{prop:rb}
Fix $H \in (0, 1/2]$, $\eta > 0$, $\xi_0 > 0$, and a compact
window $[t_0, t_1] \subset (0, 1)$. Define the local crossover
scale
\[
  s^{\star}(t) \;=\;
  \Bigl(\tfrac{c_H}{\eta H}\, t^{1 - 2H}\Bigr)^{1/(1-H)},
\]
where $c_H$ is the leading constant in
$\sqrt{\Var(Z^H_{t+s} - Z^H_t)} = c_H s^H + o(s^H)$ as
$s \downarrow 0$ (uniform in $t \in [t_0, t_1]$). Suppose, in
addition, that the persistence probability of the RL-fBm
increment process below zero has the same scaling exponent as
that of fBm of the same Hurst index, that is,
\begin{equation}
  \Prob\!\left[\,Z^H_{t+j} - Z^H_t < 0,\; j=1,\ldots,k\,\right]
  \;=\; k^{-(1-H)+o(1)} \quad (k \to \infty).
  \label{eq:hyp-rl}
\end{equation}
Then for $t \in [t_0, t_1]$ and integer $k$ with
$1 \ll k \ll T s^{\star}(t)$,
\[
  \Prob\!\left[\Tp(t) \ge k\right] \;=\; k^{-(1-H)+o(1)},
\]
and consequently, under this hypothesis, $\theta(H) = 1 - H$ on
the variance trajectory of the rough Bergomi model.
\end{proposition}

\begin{proof}[Proof sketch]
\emph{Step 1 (drift is sub-leading).}\,
Expanding $g_t(s) = \tfrac{\eta}{2}[(t+s)^{2H} - t^{2H}]$ for
$s \to 0$ at fixed $t > 0$ gives
$g_t(s) = \eta H\, t^{2H-1} s + O(s^2)$. The standard deviation
of the RL-fBm increment is $c_H s^H + o(s^H)$ (purely-new-noise
contribution $\int_t^{t+s}(t+s-u)^{2H-1}du = s^{2H}/(2H)$
dominates the memory-blend term, which is $O(s)$ for fixed
$t > 0$). The drift-to-noise ratio is therefore
\begin{equation}
  r(s, t) \;=\; \frac{g_t(s)}{c_H s^H}
    \;=\; \frac{\eta H\, t^{2H-1}}{c_H}\, s^{1 - H}
    \;+\; o(s^{1-H}),
  \label{eq:r}
\end{equation}
which vanishes as $s \downarrow 0$ for any $H < 1$ and is
bounded by a constant on $[t_0, t_1]$. Setting
$r(s^{\star}, t) = 1$ yields the crossover scale stated in the
proposition. Step 1 is rigorous and uses only the Volterra
integral representation of $Z^H$.

\emph{Step 2 (transfer of persistence exponent).}\,
By the hypothesis Eq.~\eqref{eq:hyp-rl}, the RL-fBm increment
process satisfies
$\Prob[Z^H_{t+j} - Z^H_t < 0,\; j=1,\ldots,k]
= k^{-(1-H)+o(1)}$.
Step~1 establishes that the deterministic threshold $g_t(s)$
governing the actual event Eq.~\eqref{eq:rb-event} is sub-leading
to the noise scale $c_H s^H$ in the regime
$s \ll s^{\star}(t)$. The hypothesis is therefore the only
non-rigorous input; it is discussed and motivated in
Remark~\ref{rem:rl-gap} below.

\emph{Step 3 (drift does not change the exponent).}\,
The threshold $g_t(s)$ in Eq.~\eqref{eq:rb-event} is positive,
deterministic, and sub-leading by Step~1, hence asymptotically
indistinguishable from the zero threshold on the persistence
event. By \citet[Theorem 1.3]{aurzadadereichlifshits2018},
adding a deterministic perturbation that grows slower than the
typical noise scale leaves the persistence exponent unchanged
(only the prefactor is affected). Combining Steps~1--3 yields
the claim.
\end{proof}

\begin{remark}[Status of hypothesis Eq.~\eqref{eq:hyp-rl}]
\label{rem:rl-gap}
The persistence-exponent assumption
Eq.~\eqref{eq:hyp-rl} is promoted to a hypothesis of
Proposition~\ref{prop:rb} rather than a conclusion. The
heuristic argument is that $\{Z^H_{t+s} - Z^H_t\}_{s\ge 0}$ is a
centred Gaussian process, locally self-similar of index $H$,
whose covariance asymptotes at small lags to that of a stationary
fBm of the same index; \citet{molchan1999} then governs the
persistence of the locally tangent fBm. To upgrade this heuristic
to a proof one must transfer the persistence exponent from the
locally tangent fBm to the non-stationary RL-fBm increment
process at fixed $t > 0$. Persistence of self-similar Gaussian
processes with stationary increments is treated systematically by
\citet{aurzadasimon2015}; the case of self-similar Gaussian
processes with non-stationary increments at an interior reference
time is, to our knowledge, not covered there. The missing argument
is not attempted here. The empirical match in
Figure~\ref{fig:heatmap}, where $\hat\theta(H)$ recovers $1 - H$
to two decimal places independently of $\eta$, is reported as
numerical support for the hypothesis.
\end{remark}

\begin{remark}\label{rem:rb-numeric}
For the parameters used in Figure~\ref{fig:heatmap}
($\eta = 0.3$, $\xi_0 = 0.04$, $T = 2^{16}$,
$H \in [0.05, 0.30]$), one has $s^{\star}(t) > 1$ for all
$t \in [0.1, 0.9]$ (the simulated horizon is the unit interval).
The drift-to-noise ratio of Eq.~\eqref{eq:r} satisfies
$r(s, t) \le 0.28$ uniformly over
$(s, t) \in [0, 1]\times[0.1, 0.9]$, attaining its maximum at the
endpoint $(s, t) = (1, 0.1)$. The Hill-MLE with CSN threshold
$\ell_0 \sim 20$ samples concentrates the tail estimate at scales
$s = \ell_0/T \sim 3 \times 10^{-4}$, where $r \le 10^{-3}$; the
noise-dominated regime is overwhelmingly the operative one.
\end{remark}

\paragraph{Caveat on the rough regime}
Proposition~\ref{prop:rb} is stated for $H \in (0, 1/2]$, the
regime relevant for rough volatility. For $H > 1/2$ both the
noise scale $s^H$ and the drift growth $s^{2H}$ have different
scaling exponents, and the tangency heuristic of
Remark~\ref{rem:rl-gap} must be supplemented by a separate
persistence estimate for RL-fBm; this case is not pursued here, as
the empirically observed range of volatility Hurst indices is
well below $1/2$.

\subsection{Rough Bergomi heatmap}

\begin{figure}[t]
  \centering
  \includegraphics[width=\linewidth]{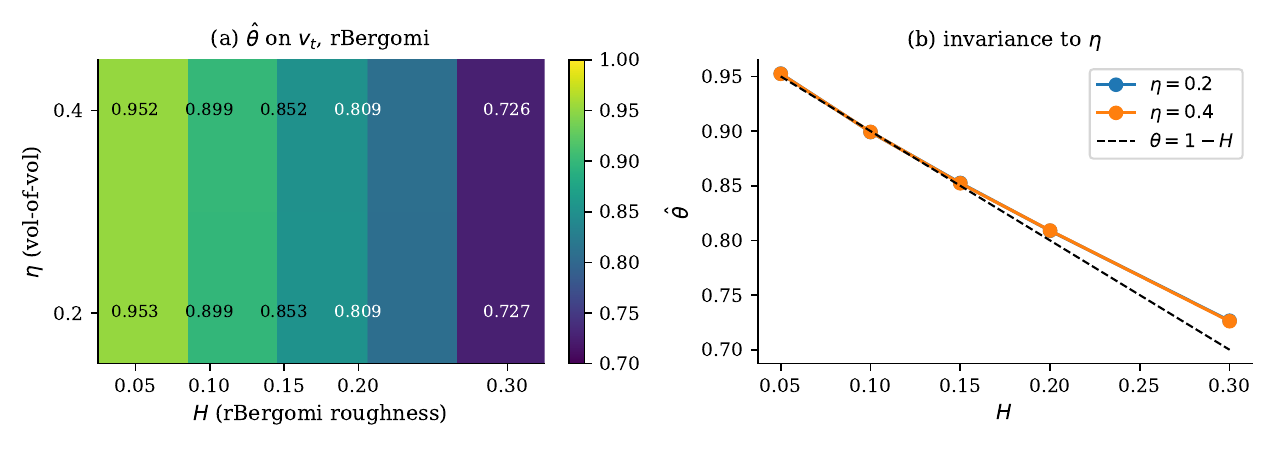}
  \caption{(a) Heatmap of $\hat\theta(H, \eta)$ for the rough
  Bergomi variance Eq.~\eqref{eq:rb}. Each cell is the mean of
  $N = 10{,}000$ paths of length $T = 2^{16}$. (b) Slices at fixed
  $\eta$ (overlapping); dashed line is $\theta = 1 - H$.}
  \label{fig:heatmap}
\end{figure}

Figure~\ref{fig:heatmap} reports $\hat\theta$ on a grid
$H \times \eta$ with $H \in \{0.05, 0.10, 0.15, 0.20, 0.30\}$ and
$\eta \in \{0.2, 0.4\}$, $\xi_0 = 0.04$, $T = 2^{16}$,
$N = 10{,}000$ paths per cell. Two observations follow:

\begin{enumerate}
  \item The estimate $\hat\theta(H, \eta)$ is essentially
  independent of $\eta$. The maximum cell-wise difference between
  $\eta = 0.2$ and $\eta = 0.4$ is $4.8\times 10^{-4}$, negligible
  on the scale of the finite-size bias.
  \item The dependence on $H$ matches
  Corollary~\ref{cor:molchan} within about $0.003$ for
  $H \le 0.15$ and within about $0.01$ for $H = 0.2$, with a
  small positive finite-size bias for $H = 0.3$, consistent with
  Table~\ref{tab:fbm}.
\end{enumerate}

These two facts taken together support $\Lp$ as a non-parametric
roughness diagnostic for rough Bergomi-type volatility, with an
$H$-dependence that is numerically stable across the tested
vol-of-vol values.

\subsection{Model comparison}
\begin{figure}[t]
  \centering
  \includegraphics[width=0.85\linewidth]{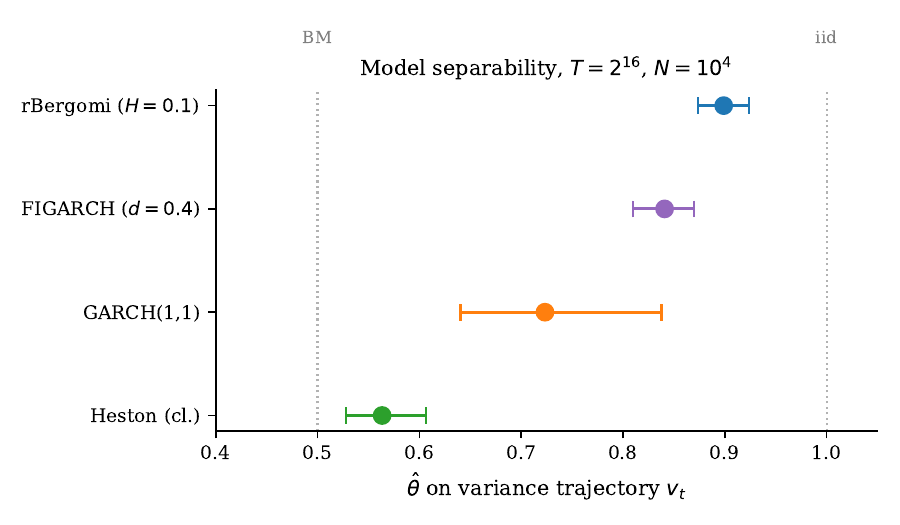}
  \caption{$\hat\theta$ on the variance trajectory $v_t$ for the
  four models, $T = 2^{16}$, $N = 10{,}000$. Dotted reference lines
  mark the BM and i.i.d.\ values from Table~\ref{tab:ctrl}.}
  \label{fig:cmp}
\end{figure}

The estimator is compared on four volatility models calibrated to
SPX-like values: rough Bergomi $H = 0.1$, $\eta = 0.3$; classical
Heston with $\kappa = 2$, $\theta_{\mathrm{var}} = 0.04$,
$\sigma = 0.5$, $\rho = -0.7$, $V_0 = 0.04$; GARCH(1,1) with
$\omega = 10^{-6}$, $\alpha = 0.05$, $\beta = 0.94$ (persistence
$0.99$); FIGARCH(1, $d$, 1) with $d = 0.4$, $\phi = 0.10$,
$\beta = 0.60$. Figure~\ref{fig:cmp} and Table~\ref{tab:cmp}
report the result.

\begin{table}[t]
\centering
\small
\caption{Model comparison, $T = 2^{16}$, $N = 10{,}000$. The 95\%
column is the cross-replicate Monte-Carlo envelope.}
\label{tab:cmp}
\begin{tabular}{@{}lccc@{}}
\toprule
model & $\hat\theta$ & SE & 95\% MC \\
\midrule
rough Bergomi $H=0.1$  & 0.8989 & 0.0001 & $[0.874, 0.924]$ \\
FIGARCH $d=0.4$        & 0.8409 & 0.0002 & $[0.810, 0.870]$ \\
GARCH$(1,1)$           & 0.7234 & 0.0007 & $[0.640, 0.838]$ \\
classical Heston       & 0.5635 & 0.0002 & $[0.528, 0.606]$ \\
\bottomrule
\end{tabular}
\end{table}

The estimator separates classical Heston from rough Bergomi by a
margin that is large compared with cross-replicate dispersion:
$\hat\theta_{\mathrm{rB}} - \hat\theta_{\mathrm{Heston}} = 0.3355$,
with combined Monte-Carlo SE $0.00024$.
GARCH(1,1) sits between the two, reflecting the burstiness of
conditional variance under high persistence
($\alpha + \beta = 0.99$). The FIGARCH value is more delicate:
$\hat\theta_{\mathrm{FIGARCH}} = 0.840$ is closer to the
rough-Bergomi value than to the GARCH value, suggesting that long
memory in volatility partially mimics path roughness at the level
of first-passage statistics. The separation between rough Bergomi
and FIGARCH is $0.0580$ (combined Monte-Carlo SE $0.00020$);
this is statistically clear but small in absolute terms, and is reported as a
methodological caveat for practitioners using $\Lp$ to
discriminate between rough and long-memory volatility.

\section{Empirical demonstration on VIX}\label{sec:vix}

The rolling estimator is applied to the daily CBOE VIX series from
the Federal Reserve Economic Data (series identifier
\texttt{VIXCLS}), covering 2000-01-03 to 2026-04-30,
$n = 6651$ trading days. VIX is used rather than realised variance
because (i) it is publicly available without licence
restrictions, and (ii) the rough-volatility literature reports a
Hurst index in the $0.05$--$0.15$ range across realised, implied,
and option-based volatility measures
\citep{gjr2014, fukasawa2022, bayerhorvathjacquier2023}. The
estimator is rank-invariant, so it is applied directly to the level
$\mathrm{VIX}_t$.

\paragraph{Scope} This deployment is intended as a public-data
stress test of the $\Lp$ statistic on an option-implied volatility
proxy, not as an estimator of the Hurst exponent of the latent
instantaneous variance. Implied and realised volatility differ by
a risk-neutral expectation and a path-dependent integral
respectively, and their fine-scale path properties need not be
identical. The analysis reports rolling-window values and a
contrast against the i.i.d.\ null; no model-specific calibration
claim is made.

\begin{figure}[t]
  \centering
  \includegraphics[width=\linewidth]{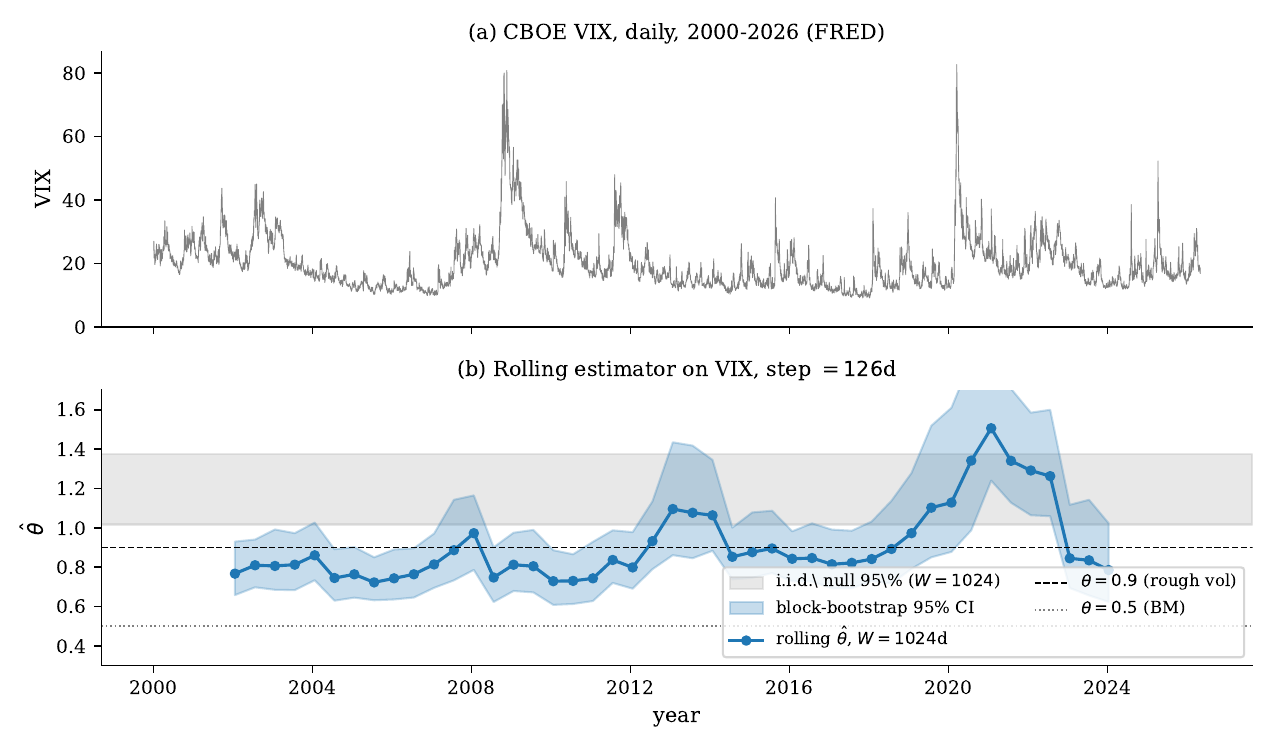}
  \caption{(a) Daily CBOE VIX from FRED, 2000--2026.
  (b) Rolling $\hat\theta$ with window 1024 days, step 126 days,
  and stationary-block-bootstrap $95\%$ confidence band.
  Horizontal lines mark the rough-volatility reference
  $\theta = 0.9$ and the Brownian-motion baseline $\theta = 0.5$.
  The horizontal grey band is the single-window 95\% empirical
  iid-null envelope at $W = 1024$.}
  \label{fig:vix}
\end{figure}

Figure~\ref{fig:vix} reports the rolling estimator with window
$W = 1024$ days (approximately four years) and step $126$ days
(approximately six months). Across the $45$ windows the mean is
$\hat\theta = 0.91$, the standard deviation is $0.19$, and the
range is $[0.72, 1.51]$. Forty-three of the forty-five windows
lie above $0.7$, consistent with $H_{\mathrm{vol}} \in (0, 0.3)$.

\paragraph{Censoring sensitivity}
Because terminal non-crossings are right-censored and can occur in
short rolling windows, Table~\ref{tab:censor} compares the
complete-case Hill estimate used in Figure~\ref{fig:vix} with the
censored-Pareto likelihood and the lower-bound treatment described
in Section~\ref{sec:est}. Including censored horizons moves the
rolling mean downward rather than upward, so the complete-case
estimate is conservative for the heavy-tail interpretation. On
average a 1024-day VIX window contains $24.4$ right-censored
horizons, of which $21.8$ lie above the selected tail threshold.

\begin{table}[t]
\centering
\small
\caption{Censoring sensitivity for the VIX rolling windows
($W=1024$, step $126$ days). Values are cross-window summaries of
$\hat\theta$.}
\label{tab:censor}
\begin{tabular}{@{}lccc@{}}
\toprule
method & mean & $\sigma$ & range \\
\midrule
complete-case Hill & 0.910 & 0.190 & [0.723, 1.506] \\
censored Pareto likelihood & 0.674 & 0.102 & [0.507, 0.985] \\
lower-bound treatment & 0.757 & 0.099 & [0.603, 1.030] \\
\bottomrule
\end{tabular}
\end{table}

\paragraph{Comparison against an empirical i.i.d.\ null}
At the single-window level, $N=1000$ i.i.d.\ uniform replicates of
length $W=1024$ give mean $1.166$, standard deviation $0.091$, and
95\% envelope $[1.017,1.376]$ (Table~\ref{tab:null}). This
single-window envelope is the grey band in Figure~\ref{fig:vix}.
To avoid treating overlapping windows as independent, $1000$
i.i.d.\ paths of length $6651$ were also simulated and evaluated
with the same rolling scheme ($W=1024$, step $126$). In the VIX data,
$35/45$ windows fall below the single-window 2.5\% null quantile
1.017. Under the overlapping-window iid null, the corresponding
count has mean $1.35$, 99\% quantile $7$, and maximum $10$ across
the $1000$ simulations; the Monte-Carlo tail probability for
observing at least $35$ such windows is $p \simeq 0.001$ with
add-one smoothing. The rolling mean is similarly extreme:
the overlapping-window iid null has mean $1.160$ and standard
deviation $0.026$, whereas VIX gives $0.910$
($p \simeq 0.001$ for a mean this small).

The deployment therefore shows a stable ordinal deviation from a
locally-i.i.d.\ benchmark on a long, real, non-stationary series.
It should not be read as proof of rough volatility for the latent
instantaneous variance; rather, it is consistent with the rough
volatility range reported in the literature
\citep{gjr2014, fukasawa2022} and demonstrates how the statistic
can be benchmarked against an exact iid null.

\section{Discussion}\label{sec:disc}

\paragraph{Main theoretical contribution} The exact
uncensored identification $\Lp = \Tp$ (Lemma~\ref{thm:Lp}) is the
load-bearing result of this paper: it converts a local HVG
statistic into a classical first-passage statistic, with two
immediate consequences. First, the i.i.d.\ null on $\Lp$ is
closed-form and parameter-free (Proposition~\ref{thm:iid}),
equivalent to the record statistic of \citet{renyi1962}. Second,
discrete-grid persistence theory gives $\theta(H) = 1 - H$ for
fBm (Corollary~\ref{cor:molchan}) and, conditionally on the
RL-fBm increment hypothesis Eq.~\eqref{eq:hyp-rl},
Proposition~\ref{prop:rb} for rough Bergomi-type volatility. The
combination yields a rank-invariant, parameter-free roughness
diagnostic with a power-law scale that is robust to the
divergence of moments.

\paragraph{Empirical contribution} On synthetic data the
estimator recovers $\theta(H) = 1 - H$ within one
cross-replicate standard deviation for $H \le 0.2$ at
$T = 2^{16}$ and resolves the positive finite-size bias for
smoother fBm paths (Table~\ref{tab:fbm}); on the rough
Bergomi heatmap it is effectively invariant to the tested vol-of-vol values
$\eta$ (Figure~\ref{fig:heatmap}); and it separates rough Bergomi
from classical Heston by $0.3355$ (combined Monte-Carlo SE
$0.00024$,
Table~\ref{tab:cmp}). On the FRED VIX series the rolling estimator
has mean $0.910$ versus an overlapping-window iid null mean
$1.160$, with Monte-Carlo $p\simeq0.001$ for both the rolling mean
and the number of windows below the single-window 2.5\% null
quantile.

\paragraph{Summary of core results} Lemma~\ref{thm:Lp}
identifies the uncensored forward visibility horizon $\Lp$ with
the classical first-passage time $\Tp$. Proposition~\ref{thm:iid}
provides the closed-form i.i.d.\ null
$\Pr[\Lp \ge k] = 1/k$ on a power-law scale.
Corollary~\ref{cor:molchan} predicts $\theta(H) = 1 - H$ for fBm
and is verified numerically in Figure~\ref{fig:fbm}.
Proposition~\ref{prop:rb} gives the corresponding rough Bergomi
prediction under an explicit persistence hypothesis. The
estimator is shown to be rank-invariant, robust to thinning and
to moderate additive noise, and to discriminate cleanly between
rough Bergomi volatility, classical Heston, GARCH(1,1) and
FIGARCH(1,$d$,1).

\paragraph{Limitations} Four points deserve emphasis. First,
the estimator inherits a finite-$T$ bias whose magnitude grows
with $H$; for the rough regime $H \le 0.2$ the bias is below
$0.02$ at $T = 2^{16}$, but for the smooth regime $H \ge 0.7$ no
length below $T = 2^{18}$ delivers a bias smaller than $0.1$.
Practitioners should report $\hat\theta$ together with the
threshold $\ell_0^{\star}$ and the tail size
$n_{\ell_0^{\star}}$. Second, the asymptotic distribution of
$\hat\theta$ under temporal dependence is, to our knowledge, not
yet established in the persistence literature; this paper uses a
stationary block bootstrap as a finite-sample surrogate. Third,
terminal right censoring can matter in short windows; in the VIX
application, including censored horizons via a censored Pareto
likelihood lowers the estimated exponent, so the complete-case
estimate is not driving the rejection of the iid benchmark. Fourth,
the separability of rough Bergomi and FIGARCH on $\Lp$ is small
(about $0.06$) and may be insufficient to falsify long-memory
volatility hypotheses on samples shorter than ten years of daily
data.

A natural extension is the multi-scale variant in which $\Lp$ is
computed under the directional-change clock of
\citet{glattfelder2011}, providing scale-localised estimates that
might further separate roughness from long memory. Beyond rough
Bergomi, the same estimator applies without modification to rough
Heston \citep{eleuchrosenbaum2019} via its Markovian lift
\citep{abijaberlarssonpulido2019, bayerbreneis2023}; whether the
$\theta(H) = 1 - H$ relation holds beyond the leading order in
such models is an open question.

\appendix

\section{Sensitivity analyses}\label{sec:appA}

\begin{figure}[t]
  \centering
  \includegraphics[width=\linewidth]{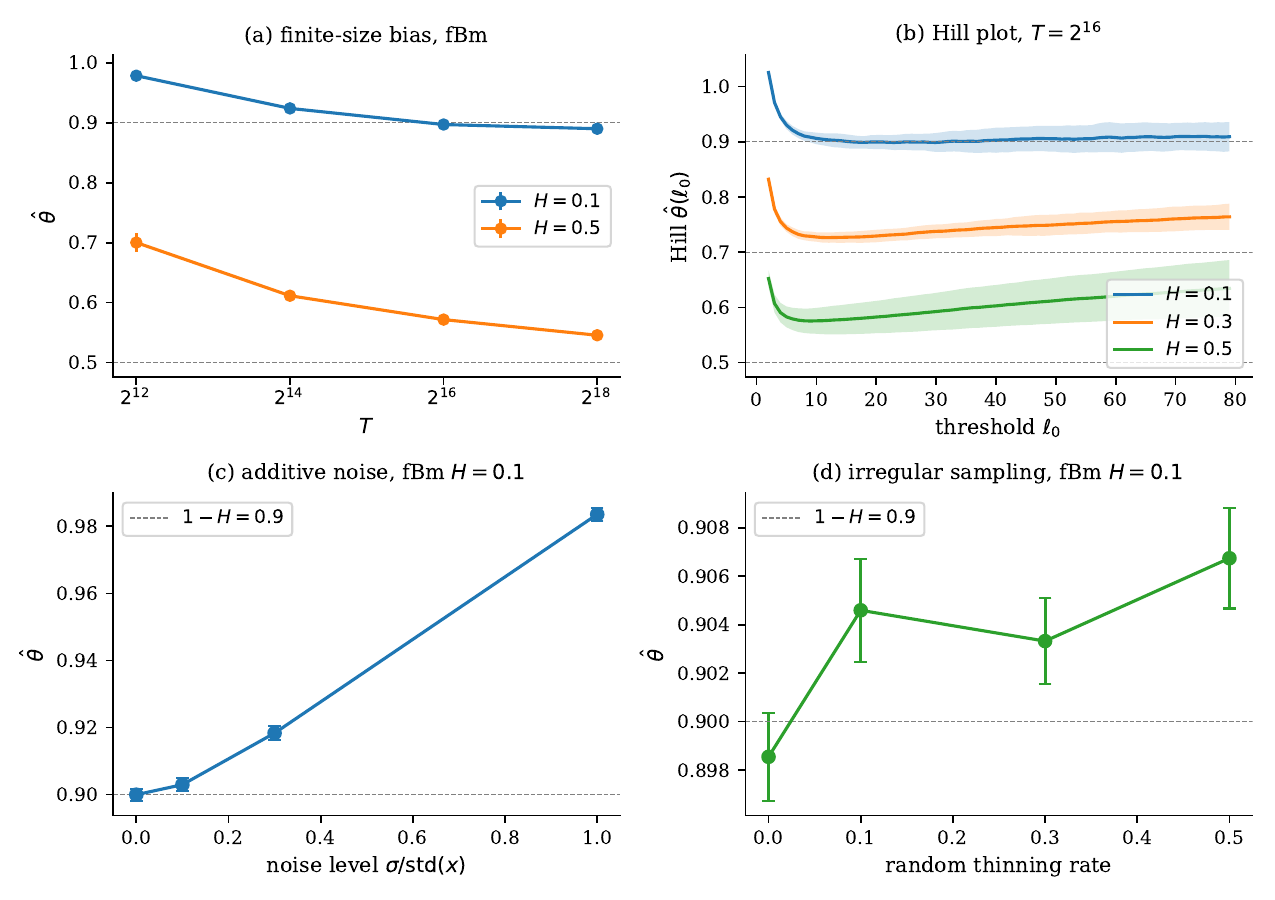}
  \caption{(a) Bias decay with $T$: $\hat\theta$ for fBm $H=0.1$
  and $H=0.5$ at $T \in \{2^{12}, 2^{14}, 2^{16}, 2^{18}\}$,
  $N = 30$ paths each; dashed lines mark $1 - H$. (b) Hill plot
  $\hat\theta(\ell_0)$ for fBm $H \in \{0.1, 0.3, 0.5\}$; bands
  are cross-replicate $\pm 1\sigma$. (c) Robustness to additive
  Gaussian noise of relative scale $\sigma / \mathrm{std}(x)$,
  fBm $H = 0.1$. (d) Robustness to random thinning at rate $r$,
  fBm $H = 0.1$.}
  \label{fig:appendix}
\end{figure}

Figure~\ref{fig:appendix}(a) shows the empirical bias decay with
$T$: for $H = 0.1$ the bias is $+0.078$ at $T = 2^{12}$ and
$-0.010$ at $T = 2^{18}$; for $H = 0.5$ the bias is $+0.200$ at
$T = 2^{12}$ and $+0.046$ at $T = 2^{18}$. Panel~(b) confirms the
existence of a wide plateau in the Hill plot: across
$\ell_0 \in [10, 60]$ the estimate moves by less than $0.02$.
Panel~(c) shows that additive Gaussian noise pushes the estimator
toward the i.i.d.\ null at rate $\hat\theta(\sigma=1.0) = 0.984$,
as expected; for moderate noise
$\sigma \le 0.3 \cdot \mathrm{std}(x)$ the bias is below $0.02$.
Panel~(d) shows that random thinning at rates up to $50\%$ leaves
$\hat\theta$ unchanged within Monte-Carlo SE; this implies that
mild irregular sampling does not contaminate the estimator.

\section{Empirical i.i.d.\ null distribution}\label{sec:appB}

To support the analysis of Section~\ref{sec:vix} and to provide
reference values for practitioners using the rolling estimator on
short windows, Table~\ref{tab:null} tabulates the empirical i.i.d.\ null
distribution of $\hat\theta$ at window lengths
$W = 2^{9}, \ldots, 2^{16}$ in Table~\ref{tab:null}. Each row
reports the cross-replicate distribution of the Hill-CSN
estimator on $N$ i.i.d.\ uniform replicates of length $W$. The
mean exceeds the asymptotic value $\theta_{\mathrm{iid}} = 1$ of
Proposition~\ref{thm:iid} by approximately $0.21$ at $W = 2^{9}$
and decays monotonically to $0.02$ at $W = 2^{16}$. The decay is
consistent with a slow rate of $\log W$ type, in line with the
standard finite-sample theory of the Hill-MLE on discrete tails
\citep[\S 4.4]{resnick2007}.

\begin{table}[!htbp]
\centering
\small
\caption{Empirical i.i.d.\ null distribution of $\hat\theta$ at
varying window length $W$. Each row is the cross-replicate
distribution of the Hill-CSN estimator on $N$ i.i.d.\ uniform
replicates of length $W$. The mean approaches the asymptotic
value $\theta_{\mathrm{iid}} = 1$ (Proposition~\ref{thm:iid})
only logarithmically slowly with $W$; practitioners using the
rolling estimator at small $W$ should consult this table for the
relevant null distribution.}
\label{tab:null}
\begin{tabular}{@{}rcccccc@{}}
\toprule
$W$ & mean & $\sigma$ & 2.5\% & 97.5\% & median & $N$ \\
\midrule
$2^{9} = 512$    & 1.211 & 0.118 & 1.010 & 1.465 & 1.198 & 1000 \\
$2^{10} = 1024$  & 1.166 & 0.091 & 1.017 & 1.376 & 1.157 & 1000 \\
$2^{11} = 2048$  & 1.136 & 0.084 & 0.994 & 1.310 & 1.130 & 1000 \\
$2^{12} = 4096$  & 1.090 & 0.056 & 0.985 & 1.201 & 1.087 & 1000 \\
$2^{13} = 8192$  & 1.060 & 0.039 & 0.988 & 1.136 & 1.058 & 1000 \\
$2^{14} = 16384$ & 1.042 & 0.025 & 0.994 & 1.092 & 1.042 & 1000 \\
$2^{15} = 32768$ & 1.030 & 0.016 & 0.999 & 1.062 & 1.030 & 500  \\
$2^{16} = 65536$ & 1.023 & 0.012 & 1.000 & 1.048 & 1.023 & 500  \\
\bottomrule
\end{tabular}
\end{table}

The bias visible at small $W$ has two complementary
interpretations. From the Hill-MLE perspective, the discrete
Pareto with exponent $\theta_{\mathrm{iid}} = 1$ is at the edge
of the regularity assumptions of \citet{hill1975}: the maximum
of $k$ i.i.d.\ values is reached approximately $\log k$ times in
the sample, so the effective tail size is logarithmically smaller
than the nominal $W$. From the CSN perspective, the threshold
$\ell_0^{\star}$ chosen by Kolmogorov--Smirnov minimisation is
constrained by the support of the data ($\ell_0^{\star} \le W$),
which biases small-$W$ estimates toward the bulk of the
distribution and away from the asymptotic regime.
Table~\ref{tab:null} quantifies the resulting bias on a grid
that spans the rolling window length $W = 1024$ used in
Section~\ref{sec:vix} as well as the full-sample length
$T = 2^{16}$ used elsewhere in the paper.

\section{Simulation details}\label{sec:appC}

\paragraph{Fractional Brownian motion} The experiments use the exact
Davies--Harte algorithm \citep{daviesharte1987} based on the
circulant embedding of the autocovariance of the fractional
Gaussian noise, with the spectral coefficients sampled via real
and imaginary Gaussians and a real-output projection.

\paragraph{Rough Bergomi} The Riemann--Liouville fBm is
simulated by the hybrid scheme of
\citet{bennedsenlundepakkanen2017} at $\kappa = 1$. The first lag
of the Volterra integral is sampled exactly via a $2 \times 2$
joint Gaussian for $(dW_k, Y_k)$ with
$Y_k = \int (t_k - s)^{H-1/2} dW_s$; the remaining lags use a
left Riemann sum at the optimal evaluation point
$b_j = ((j^{H+1/2} - (j-1)^{H+1/2})/(H+1/2))^{1/(H-1/2)}$ of
\citet[Prop.~2.8]{bennedsenlundepakkanen2017}.

\paragraph{Classical Heston} The simulations use the full-truncation Euler
scheme with reflection of the variance.

\paragraph{GARCH(1,1) and FIGARCH(1, $d$, 1)} GARCH is simulated
by the standard recursion. FIGARCH uses the truncated
infinite-order ARCH expansion at order $250$, with weights
obtained from the recursion of $(1-L)^d$ by
$\pi_j = \pi_{j-1}(j - 1 - d)/j$ and the standard ratio for
$\lambda(L) = 1 - (1 - \beta L)^{-1}(1 - \phi L)(1 - L)^d$.
Initial $250$ samples are warmed up at the unconditional
variance.


\end{document}